\documentclass{ws-ijmpe}
\usepackage[super,compress]{cite}
\usepackage{subfigure}
\usepackage{hyperref}
\hypersetup{
    colorlinks=true, 
    linktoc=all,     
    linkcolor=blue,  
    citecolor=blue
}
\def\beq{\begin{equation}}
\def\eeq{\end{equation}}
\def\be{\begin{eqnarray}}
\def\ee{\end{eqnarray}}

\def\openone{\leavevmode\hbox{\small1\kern-3.8pt\normalsize1}}
\newcommand{\gsim}{\lower.7ex\hbox{$
\;\stackrel{\textstyle>}{\sim}\;$}}
\newcommand{\lsim}{\lower.7ex\hbox{$
\;\stackrel{\textstyle<}{\sim}\;$}}

\begin{document}

\markboth{Authors' Names}{Instructions for
Typing Manuscripts (Paper's Title)}

\catchline{}{}{}{}{}

\title{TOWARDS A UNIFIED DESCRIPTION OF THE ELECTROWEAK NUCLEAR RESPONSE}

\author{\footnotesize OMAR BENHAR}

\address{INFN and Department of Physics\\
``Sapienza'' University, I-00185 Roma, Italy\\
omar.benhar@roma1.infn.it}

\author{\footnotesize ALESSANDRO LOVATO}

\address{Physics Division, Argonne National Laboratory\\
Argonne, IL 60439, USA\\
lovato@anl.gov}

\maketitle

\begin{history}
\received{Day Month Year}
\revised{Day Month Year}
\end{history}

\begin{abstract}
We briefly review the growing efforts to set up a unified framework for the description of neutrino interactions with atomic nuclei and
nuclear matter, applicable in the broad kinematical region corresponding to neutrino energies ranging between few MeV and few GeV. The emerging picture
suggests that the formalism of nuclear many-body theory can be exploited to obtain the neutrino-nucleus cross sections needed for both the 
interpretation of oscillation signals and simulation of neutrino transport in compact stars.
\end{abstract}


\ccode{PACS numbers:24.10.Cn,24.10.Lx,25.30.Pt,26.60.-c}


\section{Introduction}
\label{intro}
The description of neutrino interactions with nuclei and nuclear matter over a broad kinematical domain, besides being highly valuable in its own right,
will be required as an input for the ongoing studies aimed at addressing two
outstanding and fundamental physics issues: the violation of CP symmetry in the leptonic
sector and the mechanism leading to supernovae explosions.

Interactions of neutrinos of energy ranging from several hundreds MeV to few GeV determine the signals detected  by many experimental searches of neutrino oscillations \cite{Benhar:IJMPE_2014}, while the response of uniform nuclear matter
to weak interactions at much lower energies, of the order of few MeV, plays a critical role in simulations of neutrino transport in compact stars
\cite{Reddy:PRD,Pons:ApJ}.
In the region of high momentum transfer---typically $|{\bf q}| \gsim$ 500 MeV---in which nuclear interactions can be described within the impulse approximation (IA)
scheme \cite{benhar05}, the main difficulties involved in theoretical calculations of  the neutrino-nucleus  cross section arise from the necessity of combining a realistic model of nuclear structure and  dynamics
with a proper treatment of relativistic effects, which are known to be large. In the region of low $|{\bf q}|$, on the other
 hand, the non relativistic approximation can
be safely applied, but the scattering process may involve more than one nucleon, thus leading to the appearance of collective nuclear excitations.

Electron scattering studies have provided ample evidence that in the IA regime the approach based on the factorisation {\em ansatz} and the spectral function formalism
provides a viable computational framework to carry out accurate calculations of the nuclear inclusive cross sections.
Of critical importance, in this context, is the use a dynamical model based on a realistic nuclear hamiltonian, allowing to take into account
short range nucleon-nucleon (NN) correlations. Recently, the authors of Refs. \refcite{Lovato:2013a,Lovato:2014a} have shown that, within the approach based on correlated basis function and the cluster expansion technique, the {\em same}  hamiltonian can also be employed to obtain an effective interaction suitable for describing  long range correlations---which are known to become important in the low 
energy regime---in a fully consistent fashion.

The structure of the neutrino-nucleus cross section is outlined in Section \ref{xsec}, while Section \ref{nuclear_model} is devoted to a brief review of the
dynamical model underlying Nuclear Many-Body Theory (NMBT). The formalism employed to describe the kinematical regimes corresponding to low and
high energies are discussed in Sections \ref{nonrel} and \ref{rel}, respectively,  while in Section \ref{conclusion} we summarise the main results and state the conclusions.

\section{The lepton-nucleus cross section}
\label{xsec}

For definiteness, we shall consider charged-current neutrino-nucleus interactions. However, the formalism outlined in this section can be readily generalised
to the case of neutral current interactions  \cite{veneziano,Lovato:2014a}.

The double differential cross section of the process
\beq
\nu_\ell + A \to \ell^- + X \ ,
\eeq
where $A$ and $X$ denote the target nucleus in its ground state and the undetected nuclear final state, respectively,
can be written in the form (see, e.g., Ref. \refcite{Benhar:2006nr})
\begin{align}
\label{nucl:xsec}
\frac{d^2\sigma}{d{{\bf {\hat k}}^\prime} dk_0^\prime}& =\frac{G_F^2\,V^2_{ud}}{16\,\pi^2}\,
\frac{|\bf k^{\prime}|}{|\bf k|}\,L^{\mu\nu}\, W_{\mu\nu} \ .
\end{align}
In the above equation, $k \equiv (k_0,{\bf k})$ and $k^\prime \equiv (k^\prime_0,{\bf k}^\prime)$
are the four momenta carried by the incoming neutrino and the outgoing charged lepton, respectively,
${\bf {\hat k}}^\prime = {\bf k}^\prime/|{\bf k}^\prime|$,
$G_F$ is the Fermi coupling constant and $V_{ud}$ is the CKM matrix element coupling $u$
and $d$ quarks.

The tensor $L_{\mu\nu}$
is completely determined by lepton kinematics, whereas the nuclear tensor
$W_A^{\mu\nu}$, containing all the information on strong interaction dynamics,
describes the response of the target nucleus to weak interactions. Its definition
\begin{align}
\label{hadronictensor}
W_{\mu\nu}&= \sum_X \,\langle 0 | {J_\mu}^\dagger | X \rangle \,
      \langle X | J_\nu | 0 \rangle \;\delta^{(4)}(p_0 + q - p_X) \ ,
\end{align}
with $q\equiv(\omega,{\bf q})=k-k^\prime$, involves the initial and final states $|0\rangle$ and $|X\rangle$, with four momenta $p_0$ and $p_X$,
respectively, as well as the nuclear  current operator, $J^\mu$.

Note that the target ground state state $|0\rangle$ is independent of momentum transfer, while the state $|X\rangle$ includes at least one particle carrying momentum $\sim {\bf q}$, and the current operator   depends explicitly on ${\bf q}$. As a consequence, a fully consistent theoretical calculation of the response tensor is only possible in the kinematical regime
corresponding to $|{\bf q}|/m \ll 1$,  with $m$ being the nucleon mass, where the non relativistic approximation underlying NMBT is applicable \cite{Carlson98,LovatoSR}.

On the other hand, the treatment of the region of high momentum transfer, relevant to event analysis of many neutrino experiments, requires a theoretical approach in which the accurate description
of the nuclear ground state provided by NMBT is combined with a relativistically consistent description of  both the final state and the nuclear current.

\section{Modelling nuclear structure and dynamics}
\label{nuclear_model}

Nuclear Many-Body Theory (NMBT) is based on the tenet that nucleons can be treated as point like non relativistic particles, the
dynamics of which are described by the hamiltonian
\begin{align}
H = \sum_{i=1}^{A} \frac{{\bf p}_i^2}{2m} + \sum_{j>i=1}^{A} v_{ij}
 + \sum_{k>j>i=1}^A V_{ijk} \ .
\label{H:A}
\end{align}
In the above equation, ${\bf p}_i$ is the momentum of the $i$-th nucleon, while the potentials
$v_{ij}$ and $V_{ijk}$ describe two- and three-nucleon interactions, respectively.

The nucleon-nucleon (NN) potential
is obtained from an accurate fit to the available data on the two-nucleon system, in both bound and scattering
states, and reduces to the Yukawa one-pion-exchange potential at large distances. State-of-the-art
parametrizations of $v_{ij}$ are written in the form  \cite{Wiringa95}
\begin{align}
v_{ij}=\sum_{n=1}^{18} v^{n}(r_{ij}) O^{n}_{ij} \ , 
\label{av18:1}
\end{align}
with $r_{ij} = |{\bf r}_i - {\bf r}_j|$ and
\begin{align}
O^{n \leq 6}_{ij} = [1, (\bm{\sigma}_{i}\cdot\bm{\sigma}_{j}), S_{ij}]
\otimes[1,(\bm{\tau}_{i}\cdot\bm{\tau}_{j})]  \ ,
\label{av18:2}
\end{align}
where $\bm{\sigma}_{i}$ and $\bm{\tau}_{i}$ are Pauli matrices acting in
spin and isospin space, respectively,
and
\begin{align}
S_{ij}=\frac{3}{r_{ij}^2}
(\bm{\sigma}_{i}\cdot{\bf r}_{ij}) (\bm{\sigma}_{j}\cdot{\bf r}_{ij})
 - (\bm{\sigma}_{i}\cdot\bm{\sigma}_{j}) \ .
\end{align}
The operators corresponding to $n=7,\ldots,14$ are associated with the
non static components of the NN interaction, while those
 corresponding  to $p=15,\ldots,18$ account for small violations of charge symmetry.
Being fit to the full Nijmegen phase-shift database, as well as to
low energy scattering parameters and deuteron properties, the Argonne $v_{18}$ potential
provides an accurate description of the two-nucleon system by construction.

The inclusion of the additional three-body term, $V_{ijk}$, is needed to explain the binding energies of the
three-nucleon systems \cite{Pudliner95b}. The derivation of  $V_{ijk}$  was first discussed in the pioneering work of  Fujita and Miyazawa\cite{FujMiy},
who argued that its main contribution originates from the two-pion exchange process in which a NN interaction leads to the
excitation of one of the participating particles to a $\Delta$  resonance, which then decays in the aftermath of the interaction with a third
nucleon.

Commonly used models of the three-nucleon potential are  written in the form
\begin{align}
V_{ijk}=V_{ijk}^{2\pi}+V_{ijk}^{N} \ ,
\end{align}
where $V_{ijk}^{2\pi}$ is the attractive Fujita-Miyazawa term, while $V_{ijk}^{N}$ is a purely phenomenological repulsive term.
The parameters entering the definition of the above potential are adjusted in such a way as to reproduce the ground state energy of
the three-nucleon systems. Note that for A=3 the Schr\"odinger equation can still be solved exactly, using both deterministic or stochastic methods.

The nuclear current consists of one- and two-nucleon contributions, the latter arising from processes
in which the interaction with the beam particle involves a meson exchanged between the target nucleons.
It can be conveniently written in the form
\beq
\label{def:curr}
J^\mu = J_1^\mu + J_2^\mu = \sum_i j^\mu_i + \sum_{\rm j>i} j^\mu_{ij} \ .
\eeq
The connection between the above current operator and the nuclear hamiltonian will be discussed in the next section.

\section{Nonrelativistic regime}
\label{nonrel}

In the nonrelativistic regime, typically corresponding to $|\mathbf{q}| \lesssim 500$ MeV, both the initial and the final
state appearing in  Eq. (\ref{hadronictensor}) are eigentstates of the nonrelativistic many-body hamiltonian of Eq. (\ref{H:A}), 
satisfying the Schr\"odinger equations
\begin{equation}
\label{schroedinger}
H |0\rangle=E_0|0\rangle \ \ \ \ , \ \ \ \  H|X\rangle=E_X|X\rangle \ .
\end{equation}
In the case of light nuclear targets, the ground state wave function can be obtained from accurate stochastic approaches, such as
Green's Function Monte Carlo (GFMC) \cite{kalos:1962,grimm:1971}. In addition, 
the nuclear cross section can be conveniently rewritten in terms of the response functions $R_{\mu\nu}(q,\omega)$, obtained from Eq.\eqref{hadronictensor}
replacing the components of the current operator with their expressions obtained in the relativistic limit, taking into account terms up to order  $(|{\bf q}|/m)^2$. 

A fundamental feature of the description of neutrino-nucleus interactions at low and moderate momentum transfer 
is the possibility of employing a set of electroweak charge and current operators consistent with the  
hamiltonian of Eq.\eqref{H:A}. 

The nuclear electromagnetic current,   $J^\mu_{\rm em} \equiv (J^0_{\rm em},{\bf J}_{\rm em})$, trivially related to the vector component of  the weak current, is constrained by $H$ through the continuity equation \cite{Riska89}
\beq
\label{continuity}
{\boldsymbol \nabla} \cdot {\bf J}_{\rm em} + i [H,J^0_{\rm em}] = 0 \ .
\eeq
Note that, because the NN potential $v_{ij}$ does not commute with the charge operator $J^0_{\rm em}$, the above equation
implies that $J^\mu_{\rm em}$ involves two-nucleon contributions, as shown in Eq.\eqref{def:curr}.

The one-body electroweak operator is obtained from a non relativistic expansion of the covariant single-nucleon currents.
Two-body charge and current operators are derived within the conventional meson-exchange formalism \cite{Marcucci:2000, 
Marcucci:2005} or within the  Effective Field Theory  approach inspired to chiral perturbation theory ($\chi$EFT) \cite{Park:1993,Park:2003,Pastore:2009,Piarulli:2013,Marcucci:2013}. In this short review 
we will mainly discuss the former,  in which the dominant static part of the realistic two-nucleon potential arises from exchange 
of effective pseudoscalar  ($\pi$-like) and vector ($\rho$-like) mesons, and the corresponding charge and current operators  
 are projected out of the static components of the potential. As a consequence, the resulting vector current  is conserved by construction. 

 The conventional electroweak charge and current  operators have no free 
parameters, except the  nucleon-to-$\Delta$ axial coupling constant, which is fixed by reproducing  the tritium Gamow-Teller transition
strength in calculations based on the realistic hamiltonian discussed above.

Nonrelativistic meson-exchange currents (MEC) have been used in analyses of a variety of electromagnetic moments and 
electroweak transitions of s- and p-shell nuclei at low and intermediate values of energy and momentum transfers. Taking MEC into 
account, a good agreement has been achieved between theoretical predictions and experimental data for the $M1$ and $E2$ radiative transition rates between low-lying states 
\cite{Pervin:2007,Marcucci:2008},  $\beta$-decays and electron- and muon-capture rates \cite{Schiavilla:2002,Marcucci:2011}, 
elastic and inelastic form factors measured in $(e,e^\prime)$ scattering \cite{Wiringa:1998,Marcucci:1998,Viviani:2007,LovatoSR} 
and radiative and weak capture reactions at low energies \cite{Marcucci:2000}.

\subsection{Quantum Monte Carlo approach}
\label{GFMC}

Quantum Monte Carlo (QMC) methods were first applied to the study of properties of light nuclei over three decades ago \cite{LomnitzAdler:1981} 
(for a recent review of Quantum Monte Carlo methods for nuclear physics see, e.g., Ref. \refcite{Carlson:2014}). Within the limits of
applicability of NMBT, they provide a truly {\em ab initio} approach, allowing to perform exact calculations of a number of nuclear properties.

\subsubsection{QMC calculations of the nuclear ground state}
The first calculations employed
the variational Monte Carlo (VMC) technique, in which  the stochastic Metropolis algorithm is used for evaluating the expectation value of 
a given many-body operator using a suitably parametrized trial wave function, $\Psi_T$. The parameters entering the definition of $\Psi_T$ are optimized  
by minimizing the variational energy
\begin{equation}
E_V=\frac{\langle \Psi_T| {H} | \Psi_T \rangle }{\langle \Psi_T | \Psi_T \rangle}\geq E_0 \ , 
\end{equation}
which provides an upper bound to the ground-state energy $E_0$.
 It is worth noting that 
Monte Carlo methods can also be used in the search for the best variational parameters.

Designing an accurate variational wave function requires a deep understanding of both the structure and dynamics of the nuclear system under consideration. Standard
VMC calculations for light nuclei use a variational wave function of the form
\begin{equation}
|\Psi_T\rangle = {\mathcal{F}} |\Phi\rangle\, .
\end{equation}
The long-range behaviour is described by the Slater determinant $|\Phi\rangle$.
For example, in the case of uniform nucleon matter in the normal (i.e. non superfluid) phase,  $|\Phi\rangle$ 
is the wave function describing a non interacting Fermi gas. 
For light nuclei, on the other hand, $|\Phi\rangle$ is usually written as a sum of Slater determinants, such as those
resulting from  from a small scale shell-model calculation.

The short-range components of the wave functions are controlled by the
correlation operator $ \mathcal{F}$, the structure of which reflects the complexity
of the two- and three-nucleon potentials appearing in the nuclear hamiltonian
\begin{equation}
{\mathcal{F}}\equiv\Big(\mathcal{S}\prod_{i<j<k} {F}_{ijk}\Big)\Big(\mathcal{S}\prod_{i<j} {F}_{ij}\Big) \ .
\end{equation}
In the above equation, $\mathcal{S}$ is the symmetrisation operator, needed to fulfill the requirement of antisymmetrisation of the trial wave function,  
while  the two-body correlation operator exhibits a spin-isospin structure similar to that of the NN potential [compare to Eq.\eqref{av18:1}]
\begin{equation}
\label{old:f}
{F}_{ij}=\sum_n f^n(r_{ij}) {O}^{n}_{ij} \ , 
\end{equation}
implying that $[F_{ij},F_{ik} ] \neq 0$.
The scalar, $f^c=f^1$, and operator, $f^{n>1}$, pair correlation functions reflect the influence of the the short-distance behavior of the two-body
potential and, at the same time, satisfy the boundary conditions implied by the requirement of cluster separability. 

As discussed below (see Section \ref{CBF}), reasonably accurate correlation functions are generated
by minimising the two-body cluster contribution to the energy per particle. 
This procedure results in the derivation of eight Euler-Lagrange differential equations, involving
a set of variational parameters \cite{Wiringa:1991}. 

Three-body correlation functions are induced by both the two- and three-body potentials. As for the latter case, the form suggested 
by perturbation theory 
\begin{equation}
{F}_{ijk}=1+\sum_x \epsilon_x {V}_{ijk}(y\,r_{ij},y\,r_{jk},y\,r_{ik})\, ,
\end{equation}
is usually employed. In the above equation, the subscript $x$ labels the various contributions to the three-body force, the $\epsilon_x$ are small 
negative strength parameters, and $y$ is a scaling factor. 
 
The GFMC method \cite{kalos:1962,grimm:1971} overcomes 
the limitations of the variational wave-function by using an imaginary-time projection technique to 
enhance the ground-state component of the starting trial wave function. The method relies on the observation 
that $\Psi_T$ can be expanded in the complete set of eigenstates of the the hamiltonian according to
\begin{equation}
|\Psi_T\rangle=\sum_n c_n |n\rangle \qquad ,\qquad {H}|n\rangle = E_n |n\rangle\ ,
\end{equation}
which implies
\begin{equation}
\lim_{\tau\to\infty}e^{-({H}-E_0)\tau} |\Psi_T\rangle=c_0 |0\rangle\, ,
\end{equation}
where $\tau$ is the imaginary time. Hence,  GFMC projects out the exact lowest-energy state, provided  $\Psi_T$ it is not 
orthogonal to the true ground state, i.e. $c_0\neq 0$.

The direct calculation of $\exp[-({H} -E_0)\tau]$ for strongly-interacting
systems involves prohibitive difficulties . To circumvent this problem, the imaginary-time 
evolution is broken into $N$ small imaginary-time steps,  and complete sets of
states are inserted, in such a way that only the calculation of the short-time propagator is required. This procedure yields the expression
\begin{align}
\langle \mathbf{R}_{N+1}   | e^{-({H}-E_0)\tau}  | \mathbf{R}_1 \rangle
=& \int \langle \mathbf{R}_{N+1} | e^{-({H} -E_0)\Delta\tau} | \mathbf{R}_{N}\rangle
\langle \mathbf{R}_{N} | e^{-({H} -E_0)\Delta\tau} | \mathbf{R}_{N-1} \rangle \dots \nonumber \\
& \ \ \ \ \ \times \langle \mathbf{R}_2| e^{-({H} -E_0)} | \mathbf{R}_1 \rangle dR_2 \ldots dR_N \ , \label{eq:propagation}
\end{align}
where, for the sake of simplicity, the dependence on the spin-isospin  degrees of freedom have been omitted. 

Monte Carlo techniques are used to sample the 
paths $\mathbf{R}_i$ in the propagation. Note that, while being exact only in the  $\Delta\tau\to 0$ limit, the accuracy of Eq.\eqref{eq:propagation}
can be checked by performing several simulations with smaller time step and extrapolating to zero. 

Because nuclear interactions are strongly spin-isospin dependent, the trial wave function is written as a a sum of complex amplitudes 
for each spin-isospin state of the system
\begin{equation}
|\Psi_T\rangle = \sum_{s,t} \psi_T(\mathbf{R})\  |\chi_s(\sigma)\rangle\, |\chi_t(\tau)\rangle\, .
\end{equation}

Within standard GFMC for nuclear physics applications \cite{Carlson:1987, Pieper:2008b}, the $2^A$ many-body spin states, defined as
\begin{align}
\chi_{s=1}&= | \downarrow_1,\downarrow_2,\dots,\downarrow_A\rangle\nonumber\\
\chi_{s=2}&= | \uparrow_1,\downarrow_2,\dots,\downarrow_A\rangle\nonumber\\
&\qquad \dots\nonumber\\
\chi_{s=2^A}&= | \uparrow_1,\uparrow_2,\dots,\uparrow_A\rangle
\end{align}
are considered. The corresponding many-body isospin states can be obtained by replacing $\uparrow$ and $\downarrow$ with 
$p$ and $n$. Exploiting charge conservation, the $2^A$ isospin states can be reduced to $A!/(N!Z!)$ 
states and, by assuming that the total isospin $T$ is a good quantum number for the nucleus,
the size of $|\chi_t(\tau)\rangle$ can be further decreased.

Because the GFMC imaginary-time evolution of Eq. (\ref{eq:propagation})  involves a sum over spin and isospin
states at each step, the computing time grows exponentially with the number of particles.
The largest calculations to date have been performed for the nucleus of $^{12}$C and for the systems of 16 neutrons, corresponding to  540,672
and 65,536 spin-isospin states, respectively.

Over the past decade, the Auxiliary Field Diffusion Monte Carlo (AFDMC) has emerged as a more efficient algorithm 
for dealing  with larger nuclear system\cite{Schmidt:1999}. Within AFDMC, the spin-isospin degrees of freedom are described by 
single-particle spinors, the amplitudes of which are sampled using Monte Carlo techniques, and the 
coordinate-space diffusion in GFMC is extended to include diffusion in spin and isospin spaces.

The early applications of AFDMC were based on variational wave function containing purely central correlation functions, 
and the accuracy of this approach for systems other than pure neutron matter was limited. Recently, the authors of Ref.~\refcite{Gandolfi:2014} were able to add 
operator correlations to the trial wave function, and developed a novel  importance sampling technique, making the accuracy of AFDMC comparable 
to that  of  GFMC  even in systems containing  both protons and neutrons.

\subsubsection{QMC studies of the response functions}
\label{QMCresponse}

The calculation of the response functions 
involves major difficulties even in the region of 
$|{\bf q}| \lesssim 0.5$~GeV and $\omega$ corresponding to quasi-free kinematics, where the consequences of the nucleon's
internal structure on nuclear dynamics can be subsumed into effective many-body potentials and currents.
%
\begin{center}
\begin{figure}[h!]
\centering
\includegraphics[width=9cm]{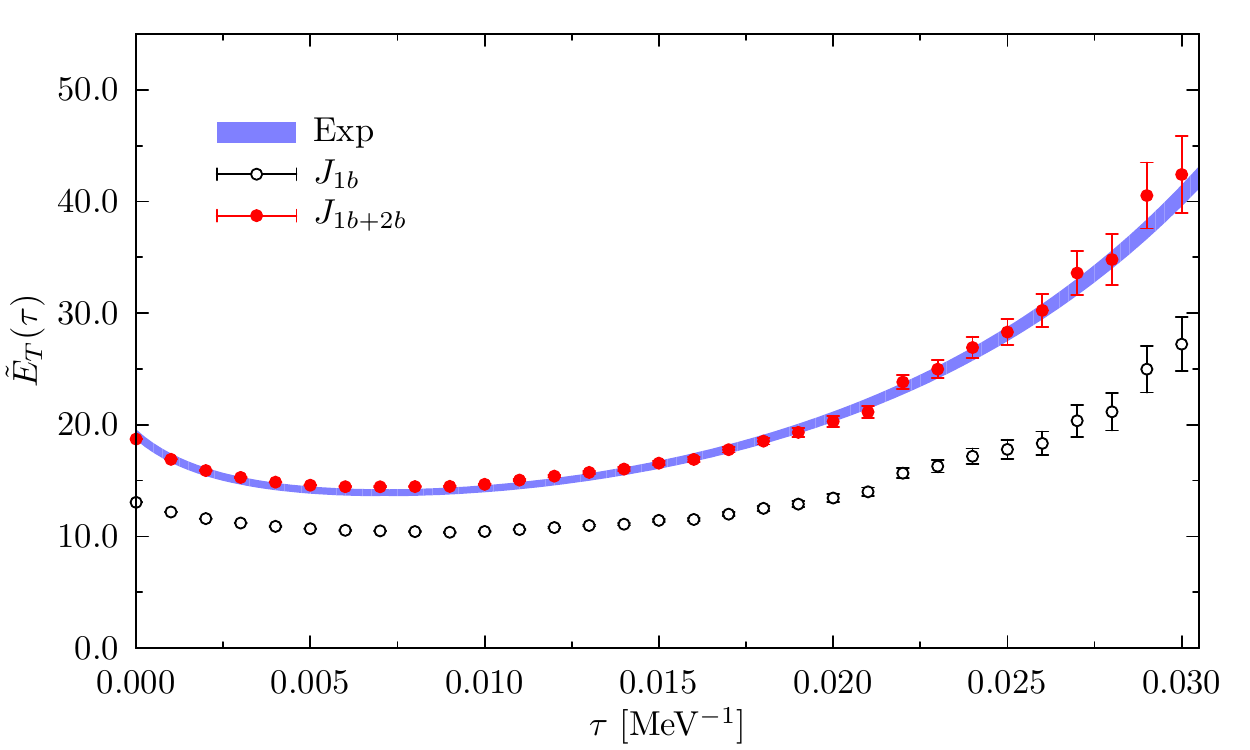}
\caption{Euclidean electromagnetic response function of $^{12}$C in the transverse channel at $|{\bf q}|=570$ MeV.  Experimental data are taken from Ref.~\protect\refcite{Jourdan:1996}.}
\label{fig:c12_em_xx_570}
\end{figure}
\end{center}
Integral properties of the response functions can be studied exploiting their sum rules, which are obtained 
from ground-state expectation values of appropriate combinations of the current operators, thus avoiding 
the calculation of the full excitation spectrum of the target nucleus. GFMC calculations of the electromagnetic sum
rules \cite{Lovato:2013}, have demonstrated that a large fraction ($\simeq 30$\%) of the strength in the 
transverse channel arises from processes involving two-body currents, and that interference effects between 
the matrix elements of one- and two-body currents play a major role \cite{Benhar:2013}. These effects are 
typically only partially, or approximately, accounted for in  existing perturbative or mean-field studies~\cite{Martini:2009,
Martini:2010,Nieves:2011,Amaro:2011}.

The main drawback of the sum rules is that they do not provide any information on the distribution of strength;
whether, for example, the calculated excess strength induced by two-body currents is mostly at large $\omega$, 
well beyond the quasi-elastic peak, or it is also found in the quasi-elastic region.  In addition, in the electromagnetic 
case, comparison of theoretical and experimental sum rules is problematic, since longitudinal and transverse response 
functions obtained from Rosenbluth separation of the measured inclusive $(e,e^\prime)$ cross sections are only available 
in the space-like region ($\omega < |{\bf q}|$) and therefore must be extrapolated into the unobserved time-like region 
($\omega > |{\bf q}|$) before ``experimental'' values for the sum rules can be determined \cite{Lovato:2013}.
\begin{center}
\begin{figure}[h]
\centering
\includegraphics[width=8.5cm]{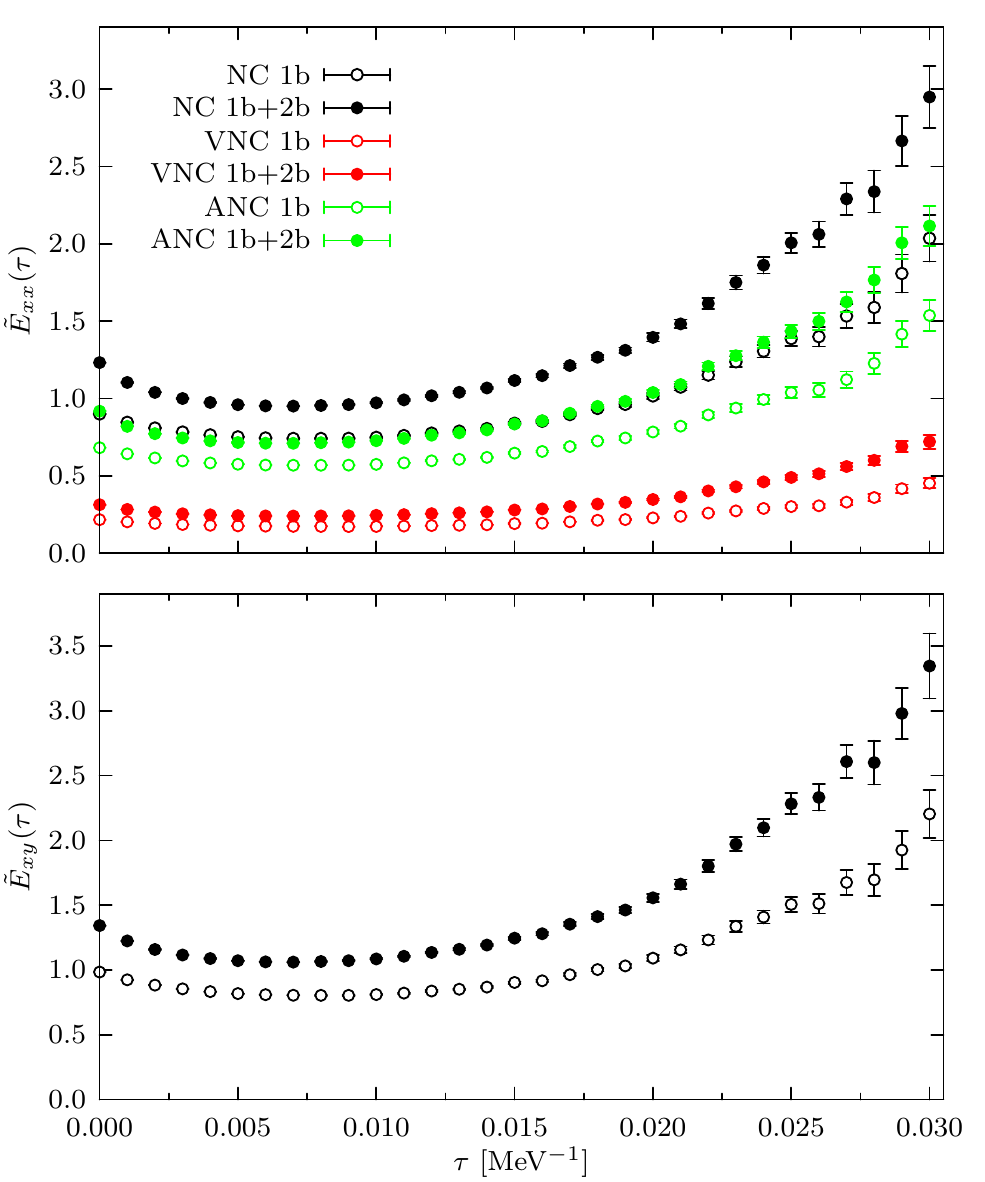}
\caption{Euclidean neutral-weak $E_{xx}$ (top panel) and $E_{xy}$ (lower panel)
response functions of $^{12}$C at $|{\bf q}| =570$ MeV.}
\label{fig:f3}
\label{fig:c12_nc_570}
\end{figure}
\end{center}
Valuable information on the $\omega$ dependence of the response functions can be inferred from the their Laplace 
transforms, also referred to as Euclidean responses\cite{Carlson:2002}, defined as
\begin{equation}
E_{\mu\nu}(q,\tau) = C_{\mu\nu}(q)\int_{\omega_{\rm th}}^\infty d\omega\,
 e^{-\tau \omega} R_{\mu\nu}(q,\omega) \ ,
\label{eq:laplace_def}
\end{equation}
where $\omega_{\rm th}$ is the inelastic threshold and the $C_{\mu\nu}$ are $q$-dependent normalization factors.
In the case of the electromagnetic longitudinal ($L$, or $\mu\nu=00$) and transverse ($T$, or $\mu\nu=xx$)
response functions, the normalization factors are~\cite{Carlson:2002} $C_L=C_T= 1/\left[G_{E}^{p}(Q^2_{\rm qe})\right]^2$, 
where $G_E^p$ is the proton electric form factor and  $Q^2_{\rm qe}=|{\bf q}|^2 - \omega_{\rm qe}^2$, 
$\omega_{\rm qe}$ being the energy transfer corresponding to 
quasi free kinematics. In the neutral-weak response functions the normalisation factors  are the same as those adopted 
in calculations of the sum rules \cite{Lovato:2013b}. 

The chief advantage of the Euclidean response is that it can be expressed as a 
ground-state expectation value
\begin{equation}
\frac{E_{\mu\nu}(|{\bf q}|,\tau)}{C_{\mu\nu}}= \frac{\langle 0| O^\dagger_{\alpha}({\bf q}) e^{-(H-E_0)\tau} 
O_{\beta}({\bf q}) |0\rangle}{\langle 0| e^{-(H-E_0)\tau}|0\rangle}\ ,
\label{eq:euc_me}
\end{equation}
where $H$ is the nuclear Hamiltonian, and $E_0$ is a trial energy controlling the 
normalisation.

In Fig.~\ref{fig:c12_em_xx_570}, taken from Ref. \refcite{Lovato:2015}, the electromagnetic transverse Euclidean response
function of $^{12}$C, $E_T$ ,  is compared to the one obtained from the analysis of the world data carried out by Jourdan~\cite{Jourdan:1996}, 
represented by the shaded band. The procedure followed to obtain the {\em experimental} Euclidean response is discussed
in Ref.~\refcite{Lovato:2015}. 
Note that, in order to emphasise the large $\tau$ behavior, the scaled Euclidean response 
$\widetilde{E}_{\mu\nu}(|{\bf q}|,\tau) = \exp[\tau\, {\bf q}^2/(2m)] E_{\mu\nu}(|{\bf q}|,\tau)$ is displayed. The results obtained by including 
only one-body or both one- and two-body terms in the electromagnetic transition operators are represented by open and solid 
circles, respectively.  Two-body current contributions substantially increase the Euclidean response over the whole range of imaginary-time, thus
implying that excess transverse strength  is generated by two-body currents not only at  $\omega \gtrsim \omega_{\rm qe}$, 
but also in the quasi-elastic and  threshold regions. The full predictions obtained including two-body currents are in excellent 
agreement with data.

A similar enhancement brought about by the two-body currents has been observed in the Euclidean neutral-weak response functions~\cite{Lovato:2015},
displayed in Fig. \ref{fig:c12_nc_570}. The neutral-current response $E_{xy}(|{\bf q}|,\tau)$ is due to the interference 
between  the vector (VNC) and axial-vector (ANC) terms of the neutral current (NC), and in the inclusive cross section the corresponding 
$R_{xy}(|{\bf q}|,\omega)$ enters with opposite sign depending on whether the process $A(\nu_l,\nu^\prime_l)$ or $A(\overline{\nu}_l,\overline{\nu}_l^{\,\prime})$
is considered~\cite{Shen:2012}. Hence, the difference between the neutrino and antineutrino cross sections turns out to be
proportional to $R_{xy}$. This difference may well have an impact on the determination of the CP-violating phase 
from neutrino and antineutrino events detected 
at DUNE \cite{LBNE:2013}.

On the other hand, since for $E_{xx}(|{\bf q}|,\tau)$ the interference between vector and  axial-vector terms vanishes, the response is simply given
by the sum of the terms with both transition operators arising from either the VNC or the ANC.
For $E_{xx}(|{\bf q}|,\tau)$ these individual contributions, along with their sum, are displayed separately.  Both the $E_{xx}(|{\bf q}|,\tau)$ and 
$E_{xy}(|{\bf q}|,\tau)$ response functions obtained retaining one-body terms only in the NC are substantially increased when two-body 
terms are also included.  This enhancement is found not only at low $\tau$, thus corroborating the sum-rule predictions of 
Ref.~\refcite{Lovato:2013b}, but in fact extends over the whole $\tau$ region.  Moreover, the individual (VNC-VNC)
and (ANC-ANC) contributions are about equally affected.

%
\begin{figure}[h!]
\centering
\includegraphics[width=8.5cm]{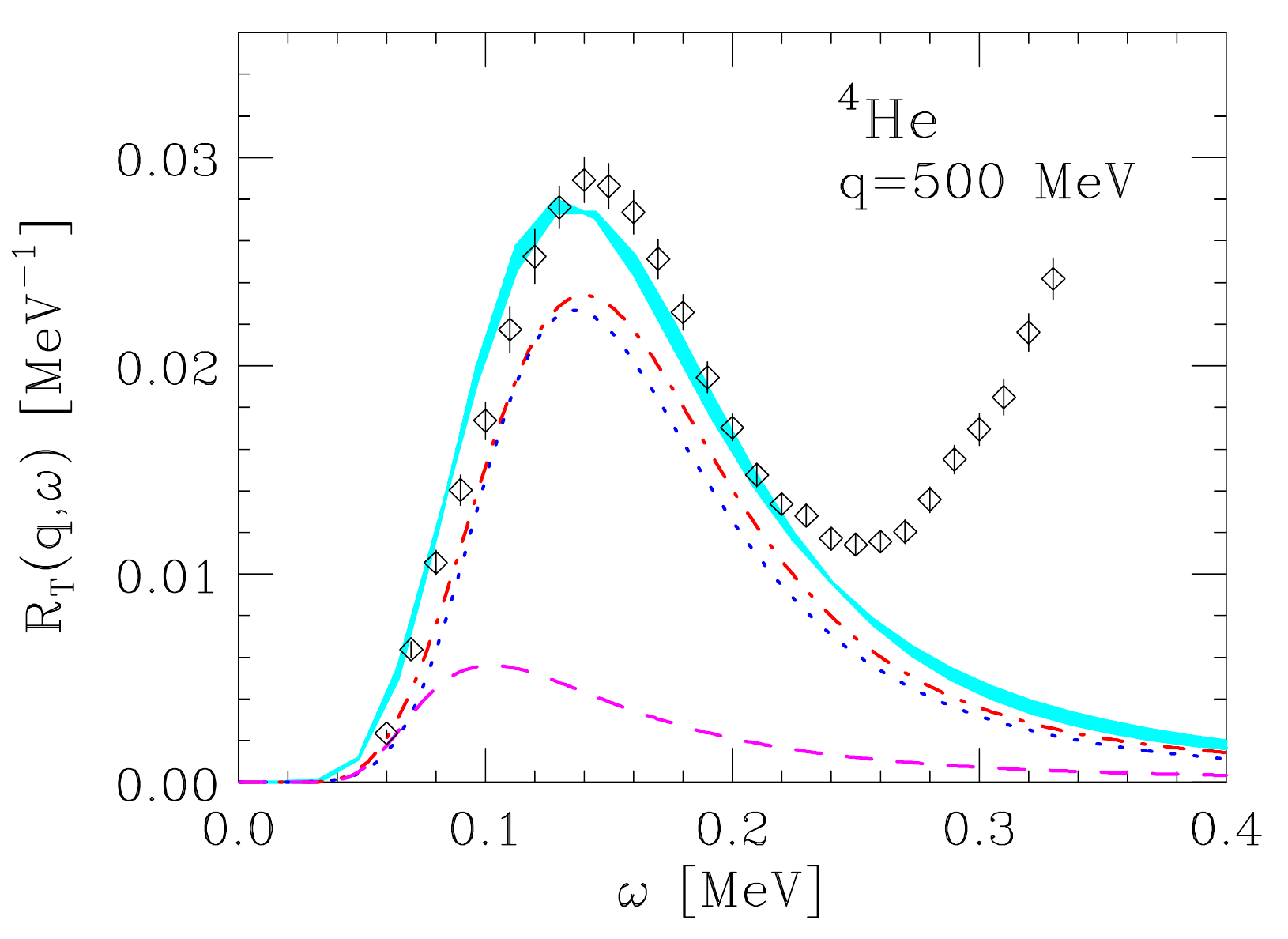}
\caption{(Color online) Transverse electromagnetic 
response functions of $^4$He at  $|{\bf q}|=500$ MeV.  Experimental data are from Ref.~\protect\refcite{Carlson:2002}.}
\label{fig:RT_inverted}
\end{figure}
%

The inversion of the Laplace transform, needed to retrieve the energy dependence of the responses, 
is long known to involve severe difficulties. A groundbreaking result has been recently reported by the
authors of Ref. \refcite{Lovato:2015}, who exploited the maximum entropy technique to obtain the 
electromagnetic longitudinal and transverse responses of $^4$He.

Figure \ref{fig:RT_inverted} shows the breakdown of the transverse response of $^4$He at $|\mathbf{q}~|=~500 \ {\rm MeV}$ 
into one-nucleon current, two-nucleon current and interference contributions\cite{Benhar:2015}. Note that the quantity 
displayed in the figure is normalized dividing by the squared proton form factor. It clearly appears that 
including the two-nucleon currents leads to a sizable enhancement of the response, and that the large positive 
contribution of the interference term peaks at energy loss $\omega < \omega_{qe}$. The agreement between
the GFMC results and the data of turns out to be remarkably good.


\subsection{Correlated Basis Functions and Custer Expansion Formalism}
\label{CBF}

The Green's Function Monte Carlo method, while providing a most powerful computational scheme to carry out {\em exact} calculations of a variety
of nuclear properties and scattering observables, is still limited to light nuclei, with $A \leq 12$.
An alternative approach, which has been extensively employed to study both medium-heavy nuclei \cite{bisconti} and nuclear matter \cite{akmal}, is based on
the use of correlated basis functions (CBF) and the cluster expansion technique (see, e.g.,  Refs. \refcite{CBF,CBF1,CBF2}). 

Let us consider, for simplicity, uniform and isospin symmetric nuclear matter. In the is case, the correlated states
are obtained from the corresponding states of the non interacting Fermi gas, $|N\rangle_{\rm FG}$, through
the transformation
\begin{align}
|N\rangle = \frac{F |N\rangle_{\rm FG} }{_{\rm FG} \langle N | F^\dagger F | N \rangle_{\rm FG}^{1/2}} \ .
\label{def:cbf}
\end{align}
The operator $F$, embodying the correlation structure induced by the NN
interaction, is written in the form (compare to Eq.\eqref{old:f})
\begin{equation}
F=\mathcal{S}\prod_{ij} F_{ij} \  .
\label{def:corrf1}
\end{equation}
The two-body correlation functions $F_{ij}$,
whose operator structure reflects the complexity of the NN potential, is written in the
form
\beq
f_{ij}=\sum_{n=1}^6 f^{n}(r_{ij}) O^{n}_{ij} \ ,
\label{def:corrf}
\eeq
including the contributions associated with the operators $O^{n\leq6}_{ij}$ of Eq.(\ref{av18:2}).

The explicit calculation of matrix elements of a many-body operator, such as the nuclear hamiltonian $H$, between correlated states
involves prohibitive difficulties, because it requires integrations over the coordinates---as well
as summations over the discrete degrees of freedom---of
many nucleons. This problem can be circumvented expanding the matrix elements
in series,  the terms of which represent the contributions of subsystems (clusters)
containing an increasing number (2, 3, \ldots, A) of particles.

Within the cluster expansion approach, the expectation value of the hamiltonian in the correlated nuclear matter ground state, the minimum of which
provides the variational estimate of the corresponding energy, can be written as \cite{CBF1}
\begin{align}
\label{def:evar}
E_V = \langle 0 | H | 0 \rangle = T_0 + \sum_n ( \Delta E_V )_n \ , 
\end{align}
where the first term in the right hand side is the energy of the non interacting Fermi gas. The contributions to the cluster expansion \eqref{def:evar} can 
be represented by diagrams and classified according to their topological structures. Selected classes of diagrams can then be summed to all orders, 
solving a set of integral equations referred to as Fermi Hyper-Netted Chain (FHNC) equations \cite{CBF1}, to obtain an accurate estimate of $E_V$

The correlation functions $f^{n}(r_{ij})$ are determined from the minimisation of the expectation value of the Hamiltonian.
The functional minimisation of the two-body cluster contribution to the energy per particle, $( \Delta E_V )_2$, leads to a set of six Euler-Lagrange 
equations, to be solved with proper constraints that force $f^c$ and $f^{(n>1)}$ to ``heal'' to one and zero, respectively. This is most efficiently achieved 
through the boundary conditions \cite{Pandharipande:1979}
\begin{align}
f^{n}(r\geq d_n) = \delta_{n1}  \ \ \ , \ \ \  \frac{df^n(r)}{dr}\mid_{d_n}= 0\, , 
\end{align}
where the healing distances $d_n$ are treated as variational parameters, to be determined from the minimisation of $E_V$.
Additional variational parameters are the quenching factors $\alpha_n$ which simulate modifications of the 
nucleon-nucleon potential, arising from the screening induced by the presence of the nuclear medium, and the set of scaling factors  $\beta_n$, often applied 
to the correlation functions $f^n$.

\subsection{The CBF effective interaction}
\label{veff}

The applications of the formalism of correlated basis function to the study of the nuclear matter response at 
low momentum transfer exploit the CBF effective interaction
\beq
\label{def:veff2}
V{^{\rm eff}} = \sum_{j>i} v^{\rm eff}_{ij} \ ,
\eeq
defined by the relation \cite{Cowell:2004}
\begin{align}
\label{def:veff}
 \langle H \rangle  = \langle 0 | H | 0 \rangle = 
T_0 \ +  \ _{\rm FG} \langle 0| V_{{\rm eff}}| 0 \rangle_{\rm FG}  \ .
\end{align}
The left hand side of the above equation, computed within the variational approach using the FHNC
summation scheme, is assumed to provide a good approximation to the ground state energy, while the 
right hand side is computed at low order of the cluster expansion, i.e. two- or three-body cluster 
level. The range of the  correlations is adjusted in such a way as to satisfy Eq. \eqref{def:veff}, which implies that the low order calculation reproduces
the variational  ground state energy.
\begin{center}
\begin{figure}[h!]
\centering
\includegraphics[width=9cm]{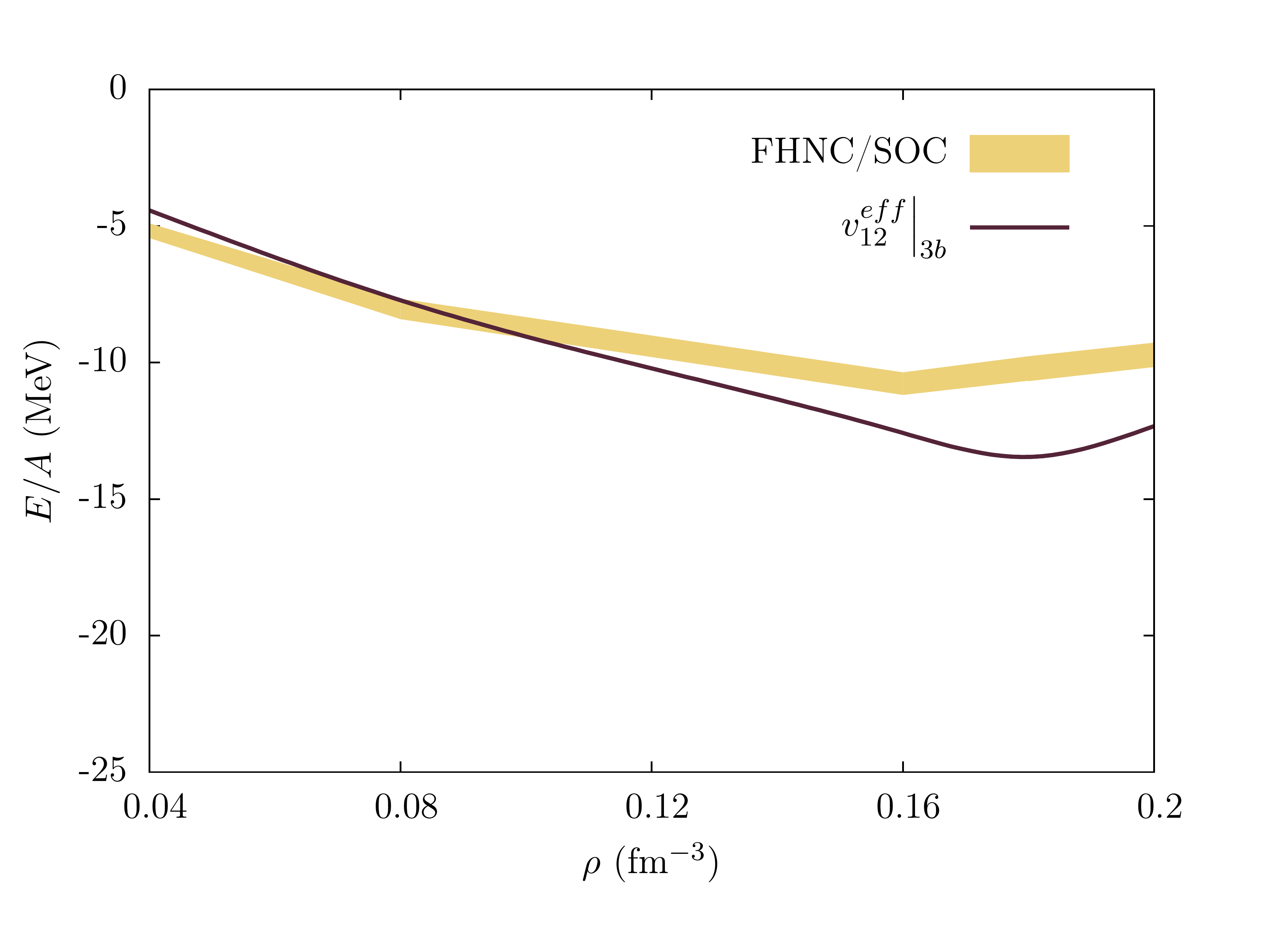}
\caption{Density dependence of the energy per particle of isospin symmetric nuclear matter in the low density regime, computed with a bare 
hamiltonian containing the Argonne $v_{6}^\prime$ potential and the UIX three-body interaction model. The solid line
displays the results obtained using the CBF effective interaction, while the full variational energy per
particle is represented by the shaded region, accounting for the uncertainty arising from the 
treatment of the kinetic energy term\cite{Lovato:2013a}. \label{fig:eos_eff}}
\label{fig:v6pUIX_eff_eos}
\end{figure}
\end{center}
The authors of Ref. \refcite{Cowell:2004}, who performed the first calculation of $V{^{\rm eff}}$ of Eq.\eqref{def:veff}, 
took into account two-body cluster diagrams only. This prescription, while leading to a very simple and transparent 
expression of $V^{\rm eff}$, fails to account for three-body forces, which are long known to play a critical 
role in  determining the energy spectrum of light nuclei, as well as the saturation properties of isospin symmetric 
nuclear matter.

The approach proposed in Ref. \refcite{Lagaris:1981}, in which interactions involving more than two nucleons 
are included through a density dependent modification of the NN potential at intermediate range, has been 
implemented in the effective interaction formalism in Refs. \refcite{Benhar:2007,Benhar:2009}. More recently, the CBF effective
interaction has been substantially improved by the authors of Refs. \refcite{Lovato:2013a,Lovato:2014a}, who 
explicitly included the three-body cluster contributions, allowing for a more realistic description of three-nucleon 
forces at microscopic level. The energy-dependence of the binding energy per nucleon in 
isospin symmetric matter obtained from the effective interaction of Refs. \refcite{Lovato:2013a,Lovato:2014a}, which 
includes the effects of the UIX model of the three-nucleon potential, is illustrated in Fig. \ref{fig:v6pUIX_eff_eos} (taken from Ref. \refcite{Lovato:2013a}). 
It turns out to be fairly close to that resulting from the full FHNC calculation, and exhibits saturation, albeit at density 
larger than the empirical value $\rho_0=0.16$ fm$^{-3}$.

\subsection{Weak response of nuclear matter at low momentum transfer}
The cooling of neutron stars is driven by the energy loss caused by the flux of neutrinos leaving the star. This effect
can be conveniently parametrized in terms of the neutrino mean free path (NMFP), which is one of the critical inputs 
required for large-scale simulations of neutrino transport~\cite{Itoh:1996}. 

Neutrino and antineutrino interactions 
in neutron matter are also relevant for understanding the evolution of the very neutron-rich matter 
formed in neutron-star mergers, since they can potentially affect the neutron to proton ratio and significantly 
impact the r-process in neutron star mergers, currently considered to be an important source for r-process 
nucleosynthesis.

The relatively low momentum scale of the above processes, typically $|{\bf q}| \lsim$ 50 MeV, allows for a 
nonrelativistic treatment of both the current operators and the final states entering the hadronic tensor.
Earlier calculations have accounted for the effects of correlations in the nuclear wave function through 
empirical effective  interactions or Landau parameters, derived from  Skyrme-like effective interactions,  and 
using the Random Phase Approximation (RPA) \cite{Iwamoto:1982,Reddy:1999}.
However, the effect of correlations on the current operators was totally neglected, as bare weak operator
were considered. It is well established that such approach is inconsistent, and that a consistent set of effective 
operators and effective interactions must be included in a more accurate treatment of the nuclear response
functions. 

The CBF effective interaction approach is best suited to define effective weak-current operators that
are consistent with the effective interaction of Eq. (\ref{def:veff}). Under the assumption that the nonrelativistic final
states entering the hadronic tensor can be described by CBF states of Eq. (\ref{def:cbf}), the effective operators can be defined
through their transition matrix elements  
\begin{align}
 \langle X | J_{\mu}^{\rm eff} | 0 \rangle = 
\frac{_{\rm FG} \langle X | \mathcal{F}^\dagger J_{\mu} \mathcal{F} | 0 \rangle_{\rm FG}}{\sqrt{_{\rm FG} \langle 0 | \mathcal{F}^\dagger \mathcal{F} | 0\rangle_{\rm FG} \, _{\rm FG}\langle X| \mathcal{F}^\dagger \mathcal{F} |X \rangle _{\rm FG} }} = \, _{\rm FG} \langle X | J_{\mu}^{\rm eff} | 0 \rangle_{\rm FG}\, .
\label{eq:cbf_1p1h}
\end{align}
Existing calculations based on the CBF effective interaction approach have only accounted for transitions between 
the correlated ground-state and correlated one particle-one hole (1p1h) excited states, which amounts to setting
$_{\rm FG} \langle X | J_{\mu}^{\rm eff} | 0 \rangle_{\rm FG} \simeq  \, _{\rm FG} \langle ph | J_{\mu}^{\rm eff} | 0 \rangle_{\rm FG}$,
where $p$ and $h$ denote both the momentum and the spin and isospin projections specifying  
single nucleon state. 

The effective operators encompass short-range correlations, but fail to account 
for long range correlations, responsible for collective modes. The $|ph\rangle_{\rm FG}$ states are not eigenstates of 
the effective hamiltonian and, as a consequence, there is a residual interaction that can induce transitions between 
different 1p1h states. 

The effect of long-range correlations can be included in the effective interaction formalism using the Tamm-Dancoff (TD)
approximation, i.e. expanding the final state in the basis of one 1p1h states according to
\begin{equation}
| X\rangle_{\rm TD} = \sum_{ph} c_{ph}^X |ph\rangle_{\rm FG}\, . 
\end{equation}
The excitation energy $E_X$ of the state $|X\rangle$, as well as the coefficients $c_{ph}^X$, are determined by solving the 
eigenvalue equation 
\begin{equation}
H_{\rm eff} | X\rangle_{\rm TD} = E_X | X\rangle_{\rm TD}\, ,
\end{equation}
with
\begin{equation} 
H^{\rm eff}=-\sum_i \frac{\nabla_{i}^2}{2m} + \sum_{j>i} v^{\rm eff}_{ij}\, .
\end{equation}
Note that the correlations defining the effective interaction used in the previous equations are the same appearing in the
definitions of the effective current operators.

\begin{figure}[h!]
\centering
\hspace*{-0.4cm}
\includegraphics[width=10.0cm,angle=0]{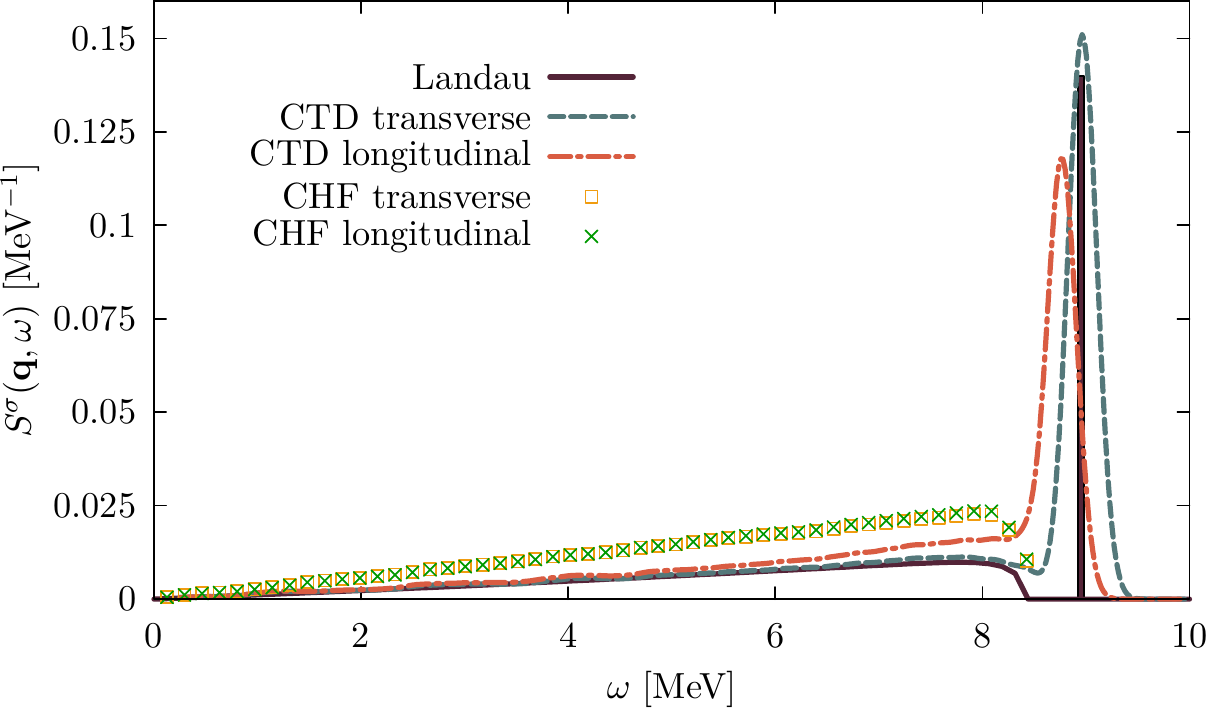}
\caption{Spin-transverse (dashed line) and spin-longitudinal (dot-dash line) responses of pure neutron 
matter, computed within the CTD and CHF approximations at $\rho = 0.16 \ {\rm fm}^{-3}$ for momentum 
transfer $|{\bf q}|=0.1\ {\rm fm}^{-1}$. The solid line has been obtained from Landau theory, according to the approach of 
Ref. \protect\refcite{Benhar:2013a} \label{fig:landau_spin}.}
\end{figure}

The effect of long-range correlation is apparent in the spin-response of pure neutron matter\cite{Lovato:2014a}, displayed in Fig. \ref{fig:landau_spin},
(taken from Ref.~\refcite{Lovato:2014a}).
The dashed and dot-dash lines represent the spin-transverse ($\mu\nu=xx+yy$) and spin-longitudinal ($\mu\nu=zz$) response 
functions of pure neutron matter at density $\rho=0.16$ fm$^{-3}$ computed in correlated Tamm-Dancoff (CTD) approximation, respectively. For comparison, 
the solid line corresponds to the spin-density response obtained from the Landau theory, the parameters of which have been 
consistently derived from the same effective interaction\cite{Benhar:2013a}.


When long-range correlations are accounted for, the peak associated with the collective excitation sticks out in both the 
spin-longitudinal and spin- transverse channels. On the other hand, in the correlated Hatree-Fock (CHF) scheme, 
in which nuclear correlations only enter via the effective operators and the quasiparticle energies, the strength of the 
response function is smoothly distributed over the particle-hole continuum.

Long-range correlations have been shown to produce a similar effect in the Fermi and Gamow-Teller responses of 
isospin-symmetric nuclear matter \cite{Cowell:2004,Benhar:2009, Lovato:2013a}, whereas no collective mode is
observed in the neutron matter density response function\cite{Lovato:2014a}.


The neutrino mean free path for low energy neutrino scattering and neutrino absorption processes in cold isospin symmetric matter has
been found to be largely affected by both short- and long-range correlations\cite{Cowell:2004}. For densities ranging
from $\rho_0/2$ to $3/2\rho_0$, the NMFP obtained from the CTD response functions is $\sim 2.5 - 3.5$ times larger than
the one of the noninteracting Fermi Gas case. In addition, the NMFP for scattering is $2$ times larger than that for 
absorption, indicating that the cross section for charged-current transitions is  $2$ times larger than the one associated
with neutral-current process.

\begin{figure}[h!]
\centering
\hspace*{-0.4cm}
\includegraphics[width=10.0cm,angle=0]{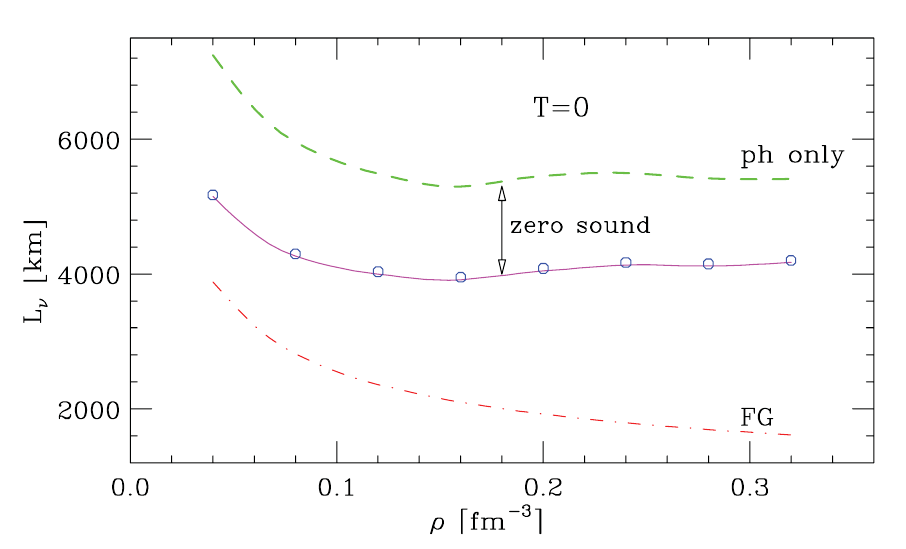}
\caption{Density dependence of the mean free path of a nondegenerate neutrino with energy $E=1$ MeV in
cold neutron matter\cite{Benhar:2013a}. Landau theory results obtained including (not including) tensor terms are represented 
by the solid  line (open circles). For comparison, Fermi gas results are also shown, by the dot-dashed line. The dashed line refers to
the NMFP obtained without accounting for the collective mode in the spin-density channel. \label{fig:landau_nmfp}}
\end{figure}

The role of long-range correlations in determining the NMFP associated with neutrino scattering processes in cold neutron matter has been investigated within
Landau theory in Ref.~\refcite{Benhar:2013a}. Figure \ref{fig:landau_nmfp}, taken from Ref. \refcite{PhysRevC.87.014601} shows the density dependence of the mean 
free path of a non  degenerate neutrino with an energy $E=1$ MeV.  The results of Landau theory, 
obtained including tensor interaction terms (solid line) and neglecting them (open circles), are compared with those corresponding to 
a free neutron gas (dot-dashed line). It clearly appears that inclusion of interaction effects leads to a large enhancement of the NMFP 
over the  whole density range. The collective mode of the spin-density response function increases the scattering cross section, 
hence reducing the NMFP, by about $25\%$. It is worth nothing that the the results of Fig. \ref{fig:landau_nmfp} have 
been confirmed by the CTD calculations reported in Ref. \refcite{Lovato:2014a}.
 
\section{Relativistic regime}
\label{rel}

The dynamical model discussed in Section \ref{nuclear_model} and the formalism of correlated basis functions
can be also employed to describe the nuclear response in the kinematical region of large momentum transfer, in which $|{\bf q}|^{-1} \ll d$, with $d$ being
the average NN separation distance in the target.  In this regime nuclear scattering can be reasonably assumed to reduce
to the incoherent sum of elementary scattering processes involving individual nucleons, and the IA is expected to be applicable.

Within the IA scenario, the difficulties associated with
the relativistic treatment of the nuclear final state and current operator are circumvented exploiting the factorisation {\em ansatz},
which amounts to: (i) neglecting the contribution of the two-nucleon current, and (ii) rewriting the nuclear final state in the form
\begin{align}
\label{factorisation}
|X \rangle = | {\bf p} \rangle \otimes |n_{\rm A-1}, {\bf p}_n \rangle \ .
\end{align}
In the above equation, the state $| {\bf p} \rangle$
describes a non interacting nucleon, while $|n_{\rm A-1}, {\bf p}_n \rangle$
is an eigenstates of the nuclear hamiltonian of Eq.\eqref{H:A}, describing the recoiling $(A-1)$-nucleon system with momentum ${\bf p}_n$.

The use of Eq. \eqref{factorisation} allows to rewrite the nuclear transition matrix element in a most simple and transparent form,
consisting of the matrix element of the one-nucleon currents between free nucleon states---which can be computed exactly, retaining
the fully relativistic expressions of the currents---and the nuclear amplitude involving the target ground state and the state
of the recoiling spectator system---which can be safely obtained from non relativistic many-body theory, since no nucleons
carrying large momenta, $\sim {\bf q}$, are involved \cite{Benhar:2015}.

The resulting expression of the differential nuclear cross section is \cite{Benhar:2006wy}
\begin{align}
\label{sigma1}
d\sigma_{IA} =  \ \sum_i \int \,d^3k \ dE \   P_i({\bf k},E) \ d\sigma_{i} \   ,
\end{align}
where $d\sigma_i$ is the corresponding cross section describing scattering on the $i$-th nucleon, the momentum and
removal energy of which are distributed according to the spectral function $P(_i{\bf k},E)$.

Accurate theoretical calculations of the
nuclear spectral function have been carried out for the few nucleon systems, with $A \leq 4$, as well as for isospin symmetric
nuclear matter (see Ref.~\refcite{Benhar:2006wy} and references therein).
For intermediate mass nuclei, realistic spectral functions have been constructed within the Local Density Approximation (LDA), in which
the results of nuclear matter calculations are combined with the empirical information extracted from the measured $(e,e^\prime p)$ cross
sections \cite{LDA}.

It is very important to realise that, owing to the presence of strong nucleon-nucleon correlations in the nuclear ground state, the recoiling
$(A-1)$-nucleon system is not necessarily left in a bound, one hole, state. Two hole-one particle states, in which one of the spectator nucleons
is excited to the continuum, typically contribute 15-20\% of the spectral function normalisation, the corresponding strength being located
at large momentum ($|{\bf k}| > 300$ MeV) and energy ($E > 40$ MeV), well outside the region corresponding to shell model states.

\begin{figure}[h!]
\centerline{\includegraphics[width=7.5cm]{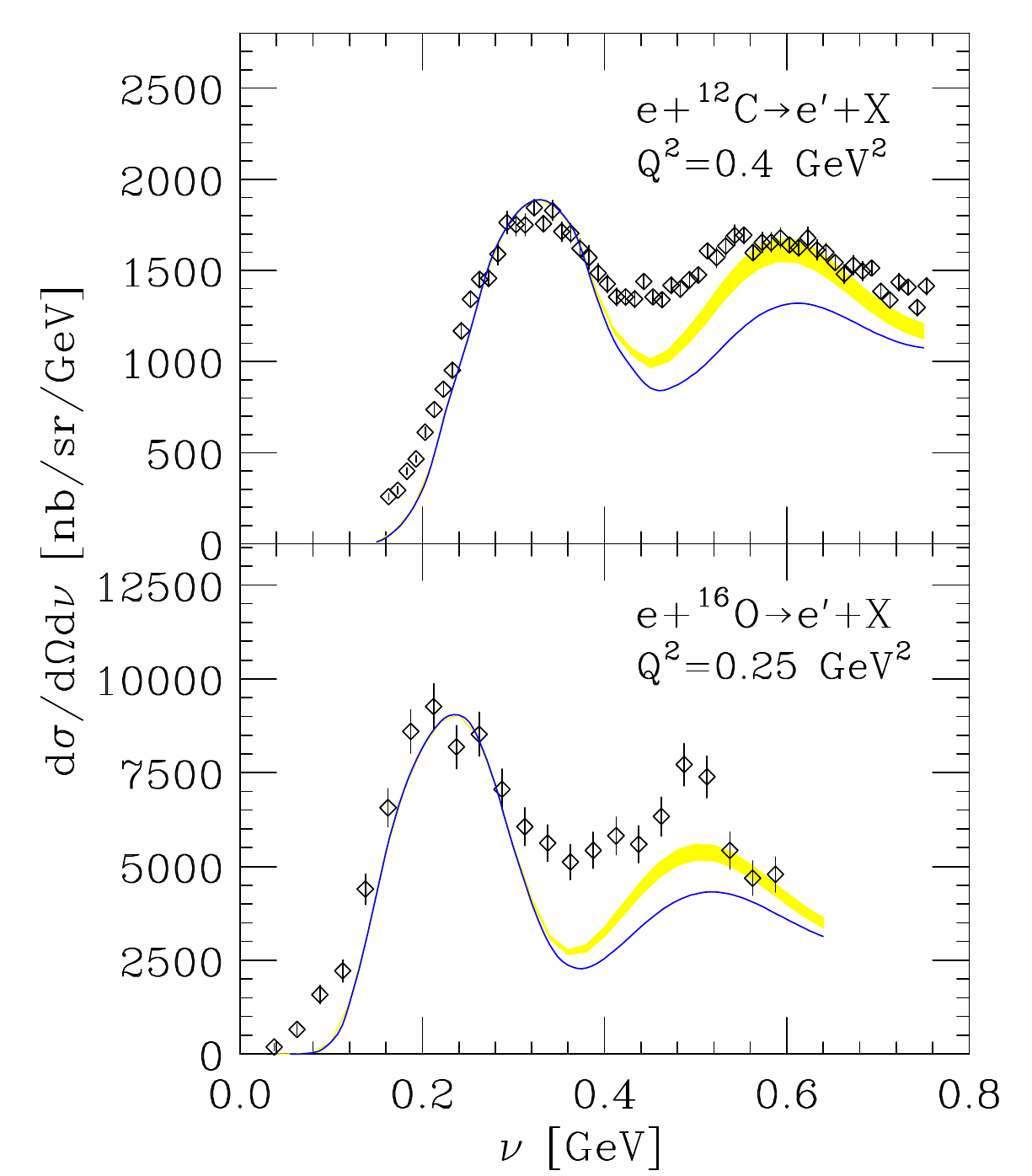}}
\caption{Top: inclusive electron scattering cross section off carbon at beam energy
$E_e = 1.3$ GeV and electron scattering angle $\theta_e = 37.5  \ {\rm deg}$, corresponding to $Q^2 = 0.4$
GeV$^2$ at the $\Delta$ production peak, as a function of the electron energy
loss $\nu$. The solid line corresponds to theoretical calculations carried out
using the proton and neutron structure functions of Ref. \protect\refcite{BR}, while
the shaded region has been obtained with those
resulting from the analysis of Ref. \protect\refcite{BenharMeloni}.
The data are taken from Ref. \protect\refcite{12C2}. Lower panel: same as in the
upper panel, but
for oxygen target, $E_e = 1.2$ GeV and $\theta_e = 32 \ {\rm deg}$, corresponding to
$Q^2 = 0.26$ GeV$^2$ at the $\Delta$ production peak. The data are taken
from Refs. \protect\refcite{Anghinolfi:95n,Anghinolfi:1996vm}.\label{ee1}}
\end{figure}

\subsection{Spectral function formalism}
\label{pke}

\begin{figure}[tbh]
\centerline{
\includegraphics[width=8.5cm]{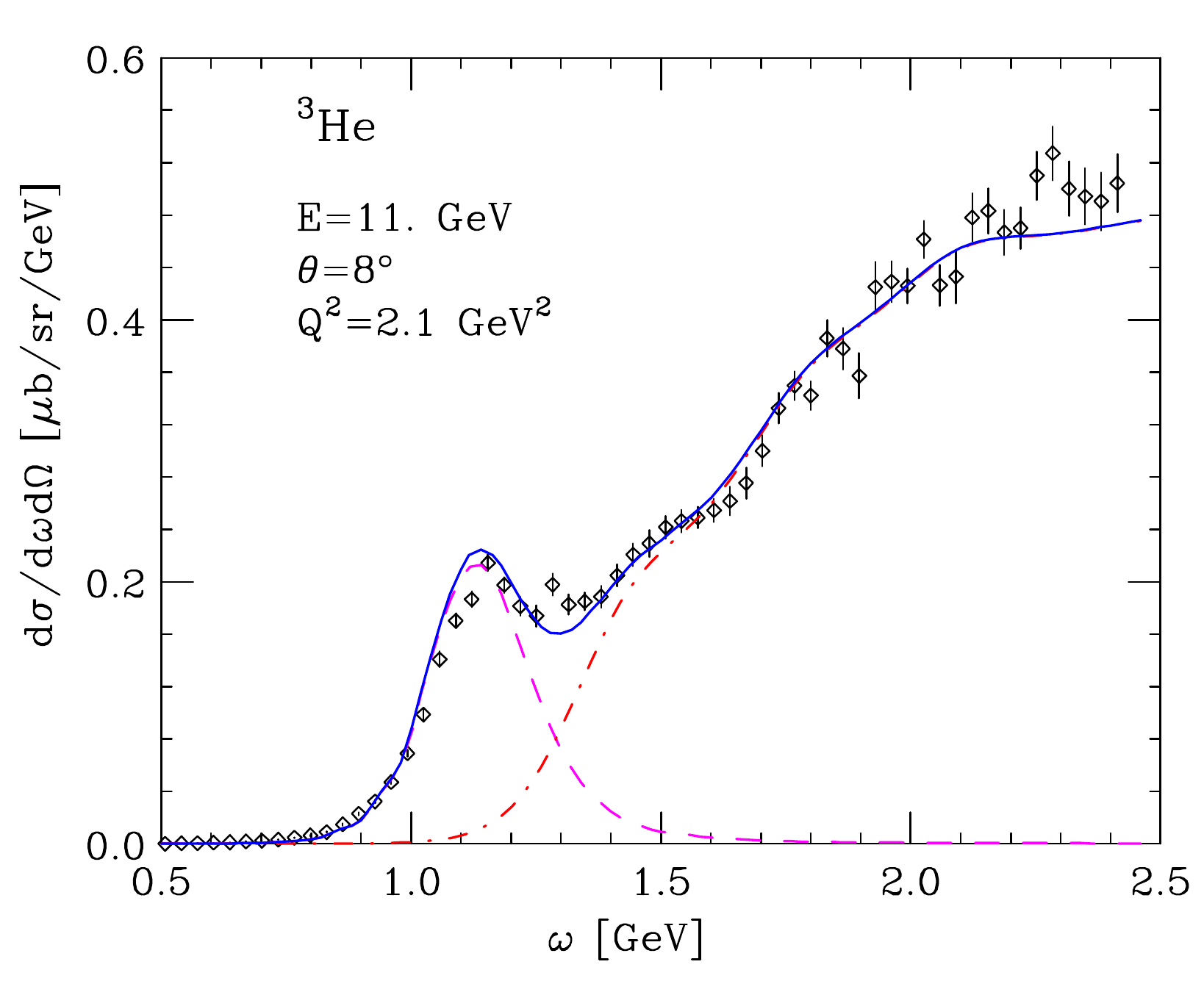}
}
\caption{Inclusive electron scattering cross section off $^3$He at beam energy
$E_e = 11$ GeV and electron scattering angle $\theta_e = 8  \ {\rm deg}$, corresponding to $Q^2 = 2.1$
GeV$^2$ at the quasi elastic peak, as a function of the electron energy
loss $\omega$. The solid line shows the full theoretical result, while the dashed and dot-dashed lines correspond to the
quasi elastic and inelastic contributions, respesctively. The data are taken from Ref. \protect\refcite{PhysRevLett.43.1143} (Adapted from Ref. \protect\refcite{BPlight}). \label{ee2}}
\end{figure}

To the extent to which the target spectral function is available, Eq. \eqref{sigma1} can be used to perform theoretical calculations of the  nuclear cross section within the IA.  The other required
input, i.e. the nucleon cross section $d\sigma_i$, can be obtained---at least in principle---from hydrogen and deuteron data. This procedure has been
widely employed to analyse the large body of inclusive electron scattering data, both in the quasi elastic sector and beyond pion production threshold.
As an example, Figs.~\ref{ee1} and \ref{ee2} show the inclusive cross section corresponding to different targets and kinematical setups.

 Inspection of Figs.~\ref{ee1} and \ref{ee2} indicates that the approach based on the IA and the spectral function formalism provides a good description of the data
 in the quasi elastic sector, corresponding to $\omega \approx \omega_{\rm QE}$, in which the elementary electron-nucleon cross section can be written in terms
 of the proton and neutron vector form factors. On the other hand, the results of Fig.~\ref{ee1} suggest that the available parametrisations of the nucleon structure functions
 in the  region in which the excitation of the $\Delta$ resonance is the dominant reaction mechanism involve a significant degree of uncertainty. At larger momentum
 transfer deep inelastic scattering clearly appears to take over at $\omega \gsim \omega_{\rm QE}$, and the agreement between theory and data over the  whole range
 of energy loss turns out to be remarkably good.
\subsection{Corrections to the impulse approximation}
\label{corrections}

The capability of obtaining accurate estimates of the nuclear cross section in the quasi elastic channel over a broad kinematical range is important for the interpretation 
of the signals detected by many neutrino experiments. For example, quasi elastic scattering provides the dominant contribution to the event sample collected by the MiniBooNE 
Collaboration using a neutrino flux of mean energy $\sim 800$ MeV\cite{AguilarArevalo:2010zc}.   

\begin{figure*}
    \subfigure{\label{fig:electrons_a}}
    \subfigure{\label{fig:electrons_b}}
    \subfigure{\label{fig:electrons_c}}
    \subfigure{\label{fig:electrons_d}}
    \subfigure{\label{fig:electrons_e}}
    \subfigure{\label{fig:electrons_f}}
    \subfigure{\label{fig:electrons_g}}
    \subfigure{\label{fig:electrons_h}}
    \subfigure{\label{fig:electrons_i}}
 \includegraphics[width=0.99\textwidth]{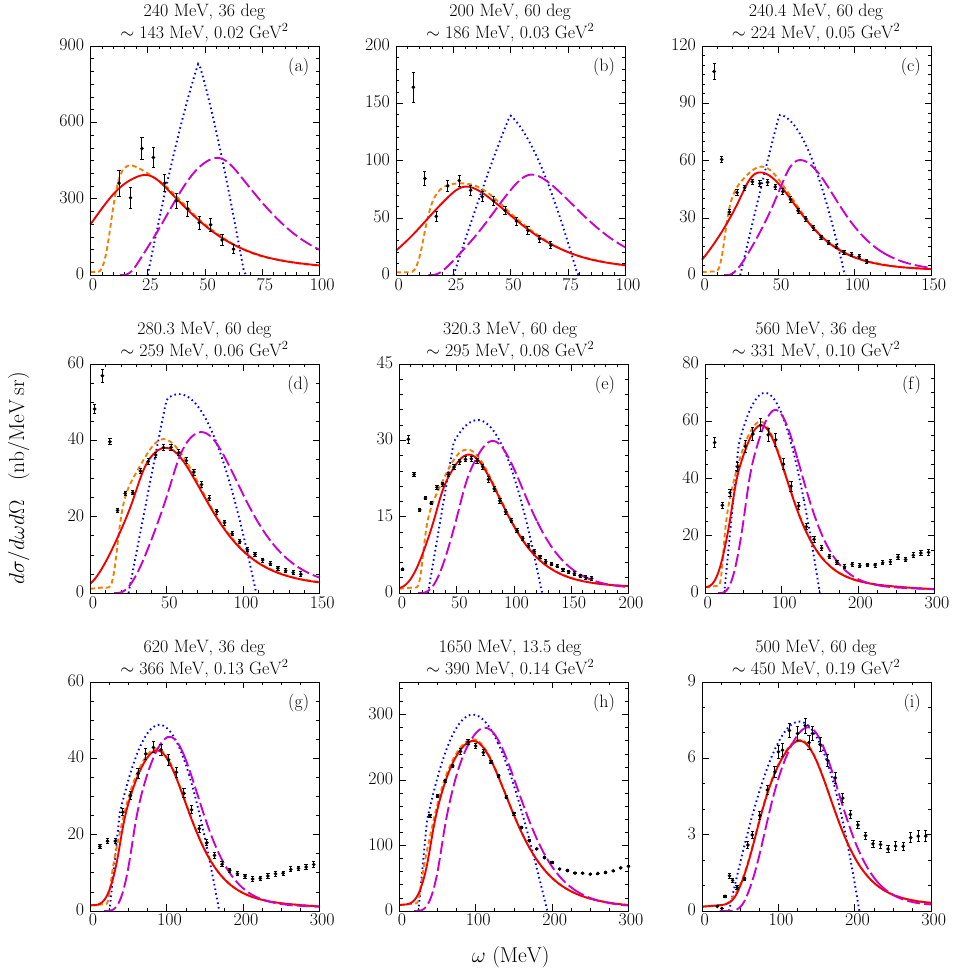}
\caption{\label{fig:electrons}(color online). Double differential electron-carbon cross sections in the QE channel of  Ref. \protect\refcite{Ankowski2013}, compared to the data of Ref. \protect\refcite{Barreau83,Baran88,Whitney74}. The solid lines correspond to the result of the full calculation, whereas the long-dashed lines have been obtained neglecting FSI.
The difference between the solid and short-dashed lines illustrates the effect of using alternative treatments of Pauli blocking. For comparison, the predictions of the Fermi gas model are
also shown, by the dotted lines. The panels are labeled according to beam energy, scattering angle, and values of $|{\bf  q}|$ and $Q^2$ at the quasielastic peak.
}
\end{figure*}

The comparison between the he results of Ref. \refcite{Ankowski2013} and the measured electron scattering cross sections (see Figure \ref{fig:electrons}),  suggests that a remarkably good agreement between theory and data in the region of the quasi elastic peak can be achieved correcting the IA results to take into account final state interactions (FSI) between the struck nucleon and the spectators, the effects of which can be  described within a generalisation of the spectral function approach, discussed in Ref. \refcite{BenharFSI}.

Additional corrections to the IA scheme arise from processes involving two-nucleon currents. It has been suggested \cite{Martini:2010,Nieves:2011yp} that the inclusion of these processes, which have been shown to play an important role in the non relativistic regime (see the discussion of Section \ref{QMCresponse}) may
in fact explain the large disagreement between  the results of Monte Carlo simulations and the double differential neutrino-carbon cross section measured by the MiniBooNE
Collaboration\cite{AguilarArevalo:2010zc}. The generalisation of the factorisation {\em ansatz} proposed in Ref. \refcite{Benhar:2013bwa,Benhar:2013} will provide a consistent framework to 
carry out calculations
of nuclear amplitudes involving two-nucleon currents, combining the fully relativistic expression of the currents and a description of nuclear dynamics taking into account 
short range correlations.

\section{Summary and perspectives}
\label{conclusion}

The results of extensive studies carried out over the past decade provide convincing evidence that the model
of nuclear structure and dynamics based on non relativistic many-body theory can be exploited to achieve a 
consistent description of neutrino interactions with nuclei---and, more generally, nuclear matter---over an energy 
range spanning three orders of magnitude. The availability of such a description will be needed to reach the level 
of accuracy required by the planned experimental searches of CP violation in the leptonic sector, as well as to 
significantly improve the modelling of neutrino transport involved in large scale simulations of compact star evolution.

The Monte Carlo approach, while being capable to yield nearly exact results, is---at least for the foreseeable future---limited to 
the quasi elastic sector and the non relativistic regime. Moreover, its extension to nuclei heavier than carbon would require
the use of enormous computing resources.
On the other hand, approximation schemes based on the same dynamical model, described by 
a realistic nuclear hamiltonian,  provide a viable alternative approach, suitable for treating  the regime of high momentum and energy 
transfer, in which relativistic effects are important and the hadronic final states involve hadrons other that protons and neutrons.

The picture emerging from the available results suggests that Monte Carlo techniques may be effectively combined with more 
approximated approaches, to both improve their accuracy and widen their scope. As an example, Monte Carlo estimates---even at variational level---of 
the nuclear amplitudes involved in spectral function calculations may 
provide valuable information, useful to reduce the theoretical uncertainty associated with the use of the 
local density approximation. In addition, the equation of state of cold nucleon matter computed using the Monte Carlo approach will provide
more precise benchmarks for the determination of the effective interactions within the formalism of correlated basis function.

The development of a unified description of neutrino interactions based on a realistic model of nuclear interactions---strongly constrained by 
phenomenology---appears to be possible, and well on its way.

\section*{Acknowledgements}

The content of this short review is partly based on a talk given by OB at the Los Alamos National Laboratory, the hospitality of 
which is gratefully acknowledged. This research is supported
by the U.S. Department of Energy, Office of Science,
Office of Nuclear Physics, under contract DE-AC02-
06CH11357 (AL). The authors are deeply indebted to Artur Ankowski, Joe Carlson, Camillo Mariani, Davide Meloni, 
Steven Pieper, Noemi Rocco, Makoto Sakuda, Rocco Schiavilla, and Robert Wiringa for countless illuminating discussions on issues 
related to the subject of this work.

\bibliographystyle{ws-ijmpe}

\bibliography{benhar}

\begin{thebibliography}{10}

\bibitem{Benhar:IJMPE_2014}
O.~Benhar, {\em Int. J. Mod. Phys. E} {\bf 58}  (2014)   013009.

\bibitem{Reddy:PRD}
S.~Reddy, M.~Prakash and J.~M. Lattimer, {\em Phys. Rev. D} {\bf 23}  (1998)
  1430005.

\bibitem{Pons:ApJ}
J.~A. Pons, S.~Reddy, J.~M. Lattimer and J.~A. Miralles, {\em ApJ} {\bf 513}
  (1999)   780.

\bibitem{benhar05}
O.~Benhar, N.~Farina, H.~Nakamura, M.~Sakuda and R.~Seki, {\em Phys. Rev. D}
  {\bf 72}  (2005)   053005.

\bibitem{Lovato:2013a}
A.~Lovato, C.~Losa and O.~Benhar, {\em Nuclear Physics A} {\bf 901}  (2013) 22
  .

\bibitem{Lovato:2014a}
A.~Lovato, O.~Benhar, S.~Gandolfi and C.~Losa, {\em Phys. Rev. C} {\bf 89}
  (2014)   025804.

\bibitem{veneziano}
O.~Benhar and G.~Veneziano, {\em Phys. Lett. B} {\bf 702}  (2011)   433.

\bibitem{Benhar:2006nr}
O.~Benhar and D.~Meloni, {\em Nucl.Phys.} {\bf A789}  (2007) 379.

\bibitem{Carlson98}
J.~Carlson and R.~Schiavilla, {\em Rev. Mod. Phys.} {\bf 70}  (1998)   743.

\bibitem{LovatoSR}
A.~Lovato {\em et~al.}, {\em Phys. Rev. Lett.} {\bf 111} (Aug 2013)   092501.

\bibitem{Wiringa95}
R.~B. Wiringa, V.~G.~J. Stoks and R.~Schiavilla, {\em Phys. Rev. C} {\bf 51}
  (1995)  ~38.

\bibitem{Pudliner95b}
B.~S. Pudliner, V.~R. Pandharipande, J.~Carlson, S.~C. Pieper and R.~B.
  Wiringa, {\em Phys. Rev. C} {\bf 56}  (1995)   1720.

\bibitem{FujMiy}
J.~Fujita and H.~Miyazawa, {\em Prog. Theor. Phys.} {\bf 17}  (1957)   360.

\bibitem{kalos:1962}
M.~H. Kalos, {\em Phys. Rev.} {\bf 128}  (1962) 1791.

\bibitem{grimm:1971}
R.~Grimm and R.~Storer, {\em Journal of Computational Physics} {\bf 7}  (1971)
  134 .

\bibitem{Riska89}
D.~O. Riska, {\em Phys. Rep.} {\bf 181}  (1989)   207.

\bibitem{Marcucci:2000}
L.~E. Marcucci, R.~Schiavilla, M.~Viviani, A.~Kievsky and S.~Rosati, {\em Phys.
  Rev. Lett.} {\bf 84}  (2000) 5959.

\bibitem{Marcucci:2005}
L.~E. Marcucci, M.~Viviani, R.~Schiavilla, A.~Kievsky and S.~Rosati, {\em Phys.
  Rev. C} {\bf 72}  (2005)   014001.

\bibitem{Park:1993}
T.-S. Park, D.-P. Min and M.~Rho, {\em Physics Reports} {\bf 233}  (1993) 341 .

\bibitem{Park:2003}
T.-S. Park, L.~E. Marcucci, R.~Schiavilla, M.~Viviani, A.~Kievsky, S.~Rosati,
  K.~Kubodera, D.-P. Min and M.~Rho, {\em Phys. Rev. C} {\bf 67}  (2003)
  055206.

\bibitem{Pastore:2009}
S.~Pastore, L.~Girlanda, R.~Schiavilla, M.~Viviani and R.~B. Wiringa, {\em
  Phys. Rev. C} {\bf 80}  (2009)   034004.

\bibitem{Piarulli:2013}
M.~Piarulli, L.~Girlanda, L.~E. Marcucci, S.~Pastore, R.~Schiavilla and
  M.~Viviani, {\em Phys. Rev. C} {\bf 87}  (2013)   014006.

\bibitem{Marcucci:2013}
L.~E. Marcucci, R.~Schiavilla and M.~Viviani, {\em Phys. Rev. Lett.} {\bf 110}
  (2013)   192503.

\bibitem{Pervin:2007}
M.~Pervin, S.~C. Pieper and R.~B. Wiringa, {\em Phys. Rev. C} {\bf 76}  (2007)
   064319.

\bibitem{Marcucci:2008}
L.~E. Marcucci, M.~Pervin, S.~C. Pieper, R.~Schiavilla and R.~B. Wiringa, {\em
  Phys. Rev. C} {\bf 78}  (2008)   065501.

\bibitem{Schiavilla:2002}
R.~Schiavilla and R.~B. Wiringa, {\em Phys. Rev. C} {\bf 65}  (2002)   054302.

\bibitem{Marcucci:2011}
L.~E. Marcucci, M.~Piarulli, M.~Viviani, L.~Girlanda, A.~Kievsky, S.~Rosati and
  R.~Schiavilla, {\em Phys. Rev. C} {\bf 83}  (2011)   014002.

\bibitem{Wiringa:1998}
R.~B. Wiringa and R.~Schiavilla, {\em Phys. Rev. Lett.} {\bf 81}  (1998) 4317.

\bibitem{Marcucci:1998}
L.~E. Marcucci, D.~O. Riska and R.~Schiavilla, {\em Phys. Rev. C} {\bf 58}
  (1998) 3069.

\bibitem{Viviani:2007}
M.~Viviani, R.~Schiavilla, B.~Kubis, R.~Lewis, L.~Girlanda, A.~Kievsky, L.~E.
  Marcucci and S.~Rosati, {\em Phys. Rev. Lett.} {\bf 99}  (2007)   112002.

\bibitem{LomnitzAdler:1981}
J.~Lomnitz-Adler, V.~Pandharipande and R.~Smith, {\em Nuclear Physics A} {\bf
  361}  (1981) 399 .

\bibitem{Carlson:2014}
J.~Carlson, S.~Gandolfi, F.~Pederiva, S.~C. Pieper, R.~Schiavilla {\em et~al.}
  (2014) \href{http://arxiv.org/abs/1412.3081}{{\ttfamily arXiv:1412.3081
  [nucl-th]}}.

\bibitem{Wiringa:1991}
R.~B. Wiringa, {\em Phys. Rev. C} {\bf 43}  (1991) 1585.

\bibitem{Carlson:1987}
J.~Carlson, {\em Phys. Rev. C} {\bf 36}  (1987) 2026.

\bibitem{Pieper:2008b}
S.~C. Pieper, {Quantum Monte Carlo Calculations of Light Nuclei}, in {\em
  {Proceedings of the "Enrico Fermi" Summer School, Course CLXIX, {\it Nuclear
  Structure far from Stability: New Physics and new Technology}}\/},  eds.
  A.~Covello, F.~Iachello, R.~A. Ricci and G.~Maino (IOS Press, Amsterdam,
  2008) p. 111.
\newblock Reprinted in La Rivista del Nuovo Cimento, {\bf 31}, 709, (2008).

\bibitem{Schmidt:1999}
K.~Schmidt and S.~Fantoni, {\em Physics Letters B} {\bf 446}  (1999) 99 .

\bibitem{Gandolfi:2014}
S.~Gandolfi, A.~Lovato, J.~Carlson and K.~E. Schmidt, {\em Phys. Rev. C} {\bf
  90}  (2014)   061306.

\bibitem{Jourdan:1996}
J.~Jourdan, {\em Nuclear Physics A} {\bf 603}  (1996) 117.

\bibitem{Lovato:2013}
A.~Lovato, S.~Gandolfi, R.~Butler, J.~Carlson, E.~Lusk, S.~C. Pieper and
  R.~Schiavilla, {\em Phys. Rev. Lett.} {\bf 111}  (2013)   092501.

\bibitem{Benhar:2013}
O.~Benhar, A.~Lovato and N.~Rocco  (2013)
  \href{http://arxiv.org/abs/1312.1210}{{\ttfamily arXiv:1312.1210 [nucl-th]}}.

\bibitem{Martini:2009}
M.~Martini, M.~Ericson, G.~Chanfray and J.~Marteau, {\em Phys. Rev. C} {\bf 80}
   (2009)   065501.

\bibitem{Martini:2010}
M.~Martini, M.~Ericson, G.~Chanfray and J.~Marteau, {\em Phys. Rev. C} {\bf 81}
   (2010)   045502.

\bibitem{Nieves:2011}
J.~Nieves, I.~Simo and M.~Vacas, {\em Phys. Rev. C} {\bf 83}  (2011)   045501.

\bibitem{Amaro:2011}
J.~Amaro, M.~Barbaro, J.~Caballero, T.~Donnelly and C.~Williamson, {\em Physics
  Letters B} {\bf 696}  (2011) 151 .

\bibitem{Carlson:2002}
J.~Carlson, J.~Jourdan, R.~Schiavilla and I.~Sick, {\em Phys. Rev. C} {\bf 65}
  (2002)   024002.

\bibitem{Lovato:2013b}
A.~Lovato, S.~Gandolfi, J.~Carlson, S.~C. Pieper and R.~Schiavilla, {\em Phys.
  Rev. Lett.} {\bf 112}  (2014)   182502.

\bibitem{Lovato:2015}
A.~Lovato, S.~Gandolfi, J.~Carlson, S.~C. Pieper and R.~Schiavilla  (2015)
  \href{http://arxiv.org/abs/1501.01981}{{\ttfamily arXiv:1501.01981
  [nucl-th]}}.

\bibitem{Shen:2012}
G.~Shen, L.~Marcucci, J.~Carlson, S.~Gandolfi and R.~Schiavilla, {\em Phys.
  Rev. C} {\bf 86}  (2012)   035503.

\bibitem{LBNE:2013}
{LBNE Collaboration}, C.~{Adams}, D.~{Adams}, T.~{Akiri}, T.~{Alion},
  K.~{Anderson}, C.~{Andreopoulos}, M.~{Andrews}, I.~{Anghel}, J.~C. {Costa dos
  Anjos} and et~al., {\em ArXiv e-prints}   (2013)
  \href{http://arxiv.org/abs/1307.7335}{{\ttfamily arXiv:1307.7335 [hep-ex]}}.

\bibitem{Benhar:2015}
O.~Benhar, A.~Lovato and N.~Rocco  (2015)
  \href{http://arxiv.org/abs/1502.00887}{{\ttfamily arXiv:1502.00887
  [nucl-th]}}.

\bibitem{bisconti}
F.~Arias~de Saavedra, C.~Bisconti, G.~C\`o and A.~Fabrocini, {\em Phys. Rep.}
  {\bf 450}  (2007)  ~1.

\bibitem{akmal}
A.~Akmal and V.~R. Pandharipande, {\em Phys. Rev. C} {\bf 56}  (1997)   2261.

\bibitem{CBF}
A.~Fabrocini and S.~Fantoni, in {\em First International Course of Condensed
  Matter Physics\/},  eds. D.~Prosperi, S.~Rosati and G.~Violini (World
  Scientific, Singapore, 1986), p.~87.

\bibitem{CBF1}
J.~W. Clark, {\em Prog. Part. Nucl. Phys.} {\bf 2}  (1979)  ~89.

\bibitem{CBF2}
S.~Fantoni and V.~R. Pandharipande, {\em Phys. Rev. C} {\bf 37}  (1988)   1697.

\bibitem{Pandharipande:1979}
V.~R. Pandharipande and R.~B. Wiringa, {\em Rev. Mod. Phys.} {\bf 51}  (1979)
  821.

\bibitem{Cowell:2004}
S.~Cowell and V.~R. Pandharipande, {\em Phys. Rev. C} {\bf 70}  (2004)
  035801.

\bibitem{Lagaris:1981}
I.~Lagaris and V.~Pandharipande, {\em Nuclear Physics A} {\bf 359}  (1981) 349
  .

\bibitem{Benhar:2007}
O.~Benhar and M.~Valli, {\em Phys. Rev. Lett.} {\bf 99}  (2007)   232501.

\bibitem{Benhar:2009}
O.~Benhar and N.~Farina, {\em Physics Letters B} {\bf 680}  (2009) 305 .

\bibitem{Itoh:1996}
N.~{Itoh}, H.~{Hayashi}, A.~{Nishikawa} and Y.~{Kohyama}, {\em The
  Astrophysical Journal Supplement Series} {\bf 102}  (1996)   411.

\bibitem{Iwamoto:1982}
N.~Iwamoto and C.~J. Pethick, {\em Phys. Rev. D} {\bf 25}  (1982) 313.

\bibitem{Reddy:1999}
S.~Reddy, M.~Prakash, J.~M. Lattimer and J.~A. Pons, {\em Phys. Rev. C} {\bf
  59}  (1999) 2888.

\bibitem{Benhar:2013a}
O.~Benhar, A.~Lovato and N.~Rocco  (2013)
  \href{http://arxiv.org/abs/arXiv:1312.1210 [nucl-th]}{{\ttfamily
  arXiv:1312.1210 [nucl-th]}}.

\bibitem{PhysRevC.87.014601}
O.~Benhar, A.~Cipollone and A.~Loreti, {\em Phys. Rev. C} {\bf 87}  (2013)
  014601.

\bibitem{Benhar:2006wy}
O.~Benhar, D.~Day and I.~Sick, {\em Rev. Mod. Phys.} {\bf 80}  (2008) 189.

\bibitem{LDA}
O.~Benhar, A.~Fabrocini, S.~Fantoni and I.~Sick, {\em Nuclear Physics A} {\bf
  579}  (1994) 493 .

\bibitem{BR}
A.~Bodek and J.~L. Ritchie, {\em Phys. Rev. D} {\bf 23}  (1981)   1070.

\bibitem{BenharMeloni}
O.~Benhar and D.~Meloni, {\em Phys. Rev. Lett.} {\bf 97}  (2006)   192301.

\bibitem{12C2}
R.~M. Sealock {\em et~al.}, {\em Phys. Rev. Lett.} {\bf 62} (Mar 1989) 1350.

\bibitem{Anghinolfi:95n}
M.~Anghinolfi {\em et~al.}, {\em Journal of Physics G: Nuclear and Particle
  Physics} {\bf 21}  (1995)  ~L9.

\bibitem{Anghinolfi:1996vm}
M.~Anghinolfi {\em et~al.}, {\em Nucl.Phys.} {\bf A602}  (1996) 405.

\bibitem{PhysRevLett.43.1143}
D.~Day {\em et~al.}, {\em Phys. Rev. Lett.} {\bf 43}  (1979) 1143.

\bibitem{BPlight}
O.~Benhar and V.~R. Pandharipande, {\em Phys. Rev. C} {\bf 47}  (1993) 2218.

\bibitem{AguilarArevalo:2010zc}
 MiniBooNE Collaboration Collaboration (A.~Aguilar-Arevalo {\em et~al.}), {\em
  Phys.Rev.} {\bf D81}  (2010)   092005.

\bibitem{Ankowski2013}
A.~M. Ankowski, O.~Benhar and M.~Sakuda, {\em Phys. Rev. D} {\bf 91}  (2015)
  033005.

\bibitem{BenharFSI}
O.~Benhar, {\em Phys. Rev. C} {\bf 87}  (2013)   024606.

\bibitem{Barreau83}
P.~Barreau {\em et~al.}, {\em Nucl. Phys. A} {\bf 402}  (1983)   515.

\bibitem{Baran88}
D.~T. Baran {\em et~al.}, {\em Phys. Rev. Lett.} {\bf 61}  (1988) 400.

\bibitem{Whitney74}
R.~R. Whitney, I.~Sick, J.~R. Ficenec, R.~D. Kephart and W.~P. Trower, {\em
  Phys. Rev. C} {\bf 9}  (1974) 2230.

\bibitem{Nieves:2011yp}
J.~Nieves, I.~Ruiz~Simo and M.~Vicente~Vacas, {\em Phys.Lett.} {\bf B707}
  (2012) 72.

\bibitem{Benhar:2013bwa}
O.~Benhar and N.~Rocco  (2013) \href{http://arxiv.org/abs/1310.3869}{{\ttfamily
  arXiv:1310.3869 [nucl-th]}}.

\end{thebibliography}

\end{document}